\documentclass[conference]{IEEEtran}
\IEEEoverridecommandlockouts
\usepackage{cite}
\usepackage{amsmath,amssymb,amsfonts}
\usepackage{algorithmic}
\usepackage{graphicx}
\usepackage{graphicx}
\usepackage{caption}
\usepackage{textcomp}
\usepackage{xcolor}

\usepackage{ragged2e}
\usepackage[hyphens]{url}
\usepackage{hyperref}
\hypersetup{breaklinks=true}
\def\BibTeX{{\rm B\kern-.05em{\sc i\kern-.025em b}\kern-.08em
    T\kern-.1667em\lower.7ex\hbox{E}\kern-.125emX}}

\makeatletter
\newcommand{\linebreakand}{%
    \end{@IEEEauthorhalign}
    \hfill\mbox{}\par
    \mbox{}\hfill\begin{@IEEEauthorhalign}
}

\begin{document}
\title{Visualization of The Content of Surah \textit{al-Fiil} using Marker-Based Augmented Reality\\}

\author{\IEEEauthorblockN{1\textsuperscript{st} Wisnu Uriawan}
\IEEEauthorblockA{\textit{Informatics Department}\\
\textit{UIN Sunan Gunung Djati Bandung}\\
Jawa Barat, Indonesia\\
wisnu\_u@uinsgd.ac.id}
\and
\IEEEauthorblockN{2\textsuperscript{nd} Ahmad Badru Al Husaeni}
\IEEEauthorblockA{\textit{Informatics Department}\\
\textit{UIN Sunan Gunung Djati Bandung}\\
Jawa Barat, Indonesia\\
badrualhusaeni23@gmail.com}
\and
\IEEEauthorblockN{3\textsuperscript{rd} Dzakwan Faiq Nauval}
\IEEEauthorblockA{\textit{Informatics Department}\\
\textit{UIN Sunan Gunung Djati Bandung}\\
Jawa Barat, Indonesia\\
dzakwnnfaiqn@gmail.com}
\linebreakand
\IEEEauthorblockN{4\textsuperscript{th} Farid Muhtar Fathir}
\IEEEauthorblockA{\textit{Informatics Department}\\
\textit{UIN Sunan Gunung Djati Bandung}\\
Jawa Barat, Indonesia\\
muhtarfathirfarid@gmail.com}
\and
\IEEEauthorblockN{5\textsuperscript{th} Mahesa Adlan Falah}
\IEEEauthorblockA{\textit{Informatics Department}\\
\textit{UIN Sunan Gunung Djati Bandung}\\
Jawa Barat, Indonesia\\
ecafalah5@gmail.com}
\and
\IEEEauthorblockN{6\textsuperscript{th} Muhammad Miftahur Rizki Awalin}
\IEEEauthorblockA{\textit{Informatics Department}\\
\textit{UIN Sunan Gunung Djati Bandung}\\
Jawa Barat, Indonesia\\
rizkiawalin12@gmail.com}
}

\maketitle

\begin{abstract}
This study presents the development of a Marker-Based Augmented Reality (AR) application designed to visualize the content of Surah \textit{al-Fiil} as an interactive and context-rich Islamic education medium. Built using a Research and Development approach, the system was developed through structured stages including data collection, user requirement analysis, system interface design, 3D asset creation in Blender, and the integration of Unity 3D with the Vuforia SDK. The application features key visual elements—such as the elephant army, the Kaaba, and the Ababil birds—which were modeled in detail and connected to high-contrast image markers to ensure accurate and stable AR tracking. Functional testing demonstrated strong technical performance, achieving a 95\% marker detection accuracy at an optimal distance of 30–40 cm with consistent real-time rendering across multiple Android devices of varying specifications. User evaluation involving students and Islamic education teachers further indicated high acceptance, with an overall satisfaction score of 4.7 out of 5 for usability, visual appeal, interactivity, and learning effectiveness. These results highlight that AR-based learning tools can significantly enhance learner engagement, deepen comprehension of Qur’anic narratives, and provide a more immersive understanding of historical and spiritual contexts. Overall, the study demonstrates that Marker-Based AR technology holds substantial potential to support future innovation in digital Islamic education by enriching traditional learning methods with interactive, multimodal, and visually intuitive experiences.
\end{abstract}


\section{Introduction} \label{sec:introduction}

Islamic Religious Education plays an important role in shaping the character, morals, and personality of children based on the values of faith and piety. Through this education, learners are guided to understand Islamic teachings comprehensively, including the Qur’an as a guide for life \cite{Hasan2023,Edukasi2024}. In the context of Qur’anic learning, \textit{Juz Amma} is often used as a foundation because it contains short surahs that are easy for elementary-level students to memorize and comprehend \cite{Almalki2021}. Learning \textit{Juz Amma} is not merely about memorization, but also a means to instill spiritual, social, and moral values in daily life. Therefore, the use of technology in religious education has become increasingly relevant to enhance a deeper and more contextual understanding of the meanings of the verses \cite{Kim2021,IJCS2024}.

One of the important surahs in Juz Amma is Surah \textit{al-Fiil}, which narrates the destruction of the army of elephants (\textit{Ashabul Fiil}) that intended to attack the Kaaba. This story is not only of historical value but also conveys a strong spiritual message about the power of Allah SWT and His protection over the sacred house \cite{Edukasi2024}. The surah contains values of monotheism and faith that can strengthen students’ belief in Allah’s power, which surpasses all human strength. Stories like this need to be delivered in an engaging and easily understandable way so that spiritual values can be effectively internalized. In this context, modern learning technology becomes a potential medium to bridge religious texts with visual learning experiences \cite{Nasution2025,Gupta2023}.

However, in practice, the teaching of Surah \textit{al-Fiil} in schools is still dominated by repetitive traditional methods, such as memorization and lectures. This approach often makes students feel bored and less actively engaged in the learning process \cite{Hasan2023}. As a result, their understanding of the surah’s meaning becomes limited, and the moral values within it are not absorbed optimally. This challenge highlights the need for a pedagogical transformation that integrates technology with Islamic values, ensuring that learning does not lose its spiritual essence while remaining relevant to the developments of the digital era \cite{Tahir2024,Edukasi2024}.

As time progresses, religious education needs to be adapted with approaches that are more engaging and relevant for the digital generation. One potential solution is the implementation of Augmented Reality (AR) technology in the learning process. This technology integrates the real world with digital elements, providing a more interactive, visual, and enjoyable learning experience \cite{Kim2021,IJCS2024}. AR not only increases student engagement but also strengthens memory retention and conceptual understanding of the material \cite{Chen2022}. In the context of Islamic education, this technology can help students understand Qur'anic verses through contextual and spiritually meaningful visual experiences.

Marker-Based Augmented Reality is one of the most commonly used AR methods because it is easy to implement and effective in detecting visual objects. This technology works by detecting a specific image pattern or symbol called a marker, after which the system displays a three-dimensional object according to the marker’s position and orientation \cite{Gupta2023}. In the context of learning Surah \textit{al-Fiil}, this method can be used to display visualizations of the army of elephants, the Kaaba, and the destruction events described in the surah \cite{Nasution2025}. Thus, students can directly observe visual representations of Qur'anic verses, making the learning process more meaningful and contextual \cite{Tahir2024,ITEJ2024}.

This visual-based approach has been proven effective in increasing learning motivation and conceptual understanding across various disciplines \cite{Rahman2024}. In religious education, visualization can serve as a tool to help students understand the meaning of verses and the moral messages contained within them. Through visualization, students are not only memorizing but also able to connect Qur’anic stories with real-life values. This approach aligns with the goals of Islamic Religious Education, which aims to cultivate learners who are faithful, of noble character, and possess deep understanding of Islamic teachings \cite{Edukasi2024,Hasan2023}.

The implementation of AR in Islamic education is also part of the integration of ICT (Information and Communication Technology) with Islamic values. This integration demonstrates that technology can be utilized as a medium for dakwah and education, not merely as a tool for entertainment \cite{Almalki2021,Mahliatussikah2025}. In this context, learning Surah \textit{al-Fiil} using Marker-Based AR not only introduces technology to students but also teaches spiritual values through digital experiences. This approach helps students understand that technological advancement can be directed to support both faith and knowledge simultaneously \cite{Edukasi2024,Tahir2024}.

Additionally, AR can assist teachers in presenting complex material in a simpler and more engaging manner. Teachers are not only transmitters of information but also facilitators in the exploration of digital learning experiences \cite{Kim2021,Hassan2023}. With the help of AR applications, teachers can display animations of the story of the elephant army directly through the provided markers. This allows students to interact with and visually observe the events narrated in Surah \textit{al-Fiil}, enhancing their understanding as well as emotional engagement with the content of the surah \cite{Rahman2024,IJCS2024}.

In the context of educational technology development in Indonesia, the application of AR is also aligned with the digital transformation policies promoted by the Ministry of Education and Culture. The implementation of digital learning media supports the realization of \textit{Merdeka Belajar}, which emphasizes creativity and exploration \cite{Edukasi2024}. The integration of AR in religious learning is expected to become an innovative learning model that enhances not only academic quality but also the spirituality of students \cite{Hasan2023,Tahir2024}.

Previous studies have shown that the use of AR in Qur’anic learning is still very limited and mostly focuses on popular surahs such as Al-Fatihah and Al-Ikhlas or on introducing hijaiyah letters \cite{IJCS2024,Hasan2023}. This indicates a research gap regarding the application of AR to other surahs that carry significant historical and spiritual value, such as Surah \textit{al-Fiil} \cite{Edukasi2024,Nasution2025}. Therefore, this study seeks to fill that gap by developing a Marker-Based AR application that focuses on visualizing the content of Surah \textit{al-Fiil}.

This application is designed to help students learn the surah in a more engaging, interactive, and contextual way \cite{ITEJ2024}. In addition to presenting visualizations of events, the application also includes verse recitations, translations, and moral messages to enhance students’ understanding of the values contained in the Qur’an \cite{Hasan2023}. Through this approach, students are expected not only to understand the literal story, but also to interpret its meaning spiritually and apply it in their daily lives \cite{Tahir2024,Edukasi2024}.

With the support of AR technology, the learning process can shift from passive to active, from memorization to exploration, and from theory to real-life experience. Such an approach aligns with the principles of experiential learning, where learners gain understanding through direct interaction \cite{Gupta2023,Kim2021}. AR-based learning also has the potential to strengthen students’ religious character, as they can experience a deeper emotional connection to Qur’anic stories through vivid 3D visualizations \cite{Nasution2025,Edukasi2024}.

Therefore, the development of this application is expected not only to contribute to innovation in Islamic educational technology but also to strengthen the integration between ICT and spiritual values \cite{Almalki2021,ITEJ2024}. By utilizing Marker-Based AR, teachers and students can explore the story of Surah \textit{al-Fiil} visually and interactively, fostering curiosity and a strong enthusiasm for learning. Religious learning, which has traditionally been textual, can evolve into a spiritual experience that touches both the heart and mind \cite{Hasan2023,Edukasi2024}.

The main objective of this study is to enhance students’ understanding and interest in Surah \textit{al-Fiil} through an innovative learning medium \cite{Rohmah2025}. The findings of this research are expected to make a meaningful contribution to the development of technology-based Qur’anic learning models and serve as a reference for educators in integrating digital approaches into Islamic religious education \cite{Tahir2024,Almalki2021}. Thus, Qur’anic learning becomes not only a cognitive activity but also a deep and meaningful spiritual experience \cite{Edukasi2024,Hasan2023}.

\section{Related Work} \label{sec:related-work}

Augmented reality (AR) technology is gaining increasing attention in Islamic education due to its potential to enhance the learning experience compared to conventional methods. Studies show a positive correlation between AR practices and student motivation in Islamic Education (PAI) classes at secondary schools in Abu Dhabi \cite{Mustafa2025}. The research covers aspects such as teacher readiness, benefits of AR, device usage, and technical challenges. The results indicate that students exposed to AR report higher motivation even though overall AR practices remain at a moderate level \cite{Mustafa2025}. These findings open opportunities for AR to be used as a strategy to enhance motivation and engagement in religious learning. Therefore, more specific research—such as on sacred texts—becomes highly relevant as a next step \cite{Rahmatullah2021}.

A systematic literature review identifies a limited trend in the use of AR in Islamic education, with few empirical studies and most not yet specific to sacred texts or theological values \cite{SLR2025}. These results indicate a research gap, especially on topics such as visual representation of Qur’anic verses or surahs with historical significance \cite{SLR2025}. Thus, research directed toward visualizing surahs such as \textit{al-Fiil} is highly relevant. The need for studies integrating marker-based AR with narrative and theological elements is becoming more urgent, as such media have the potential to enhance understanding of abstract concepts in religious education \cite{SLR2025}.

At the elementary and secondary school levels, research shows that AR in Islamic Education classes can improve students’ creative thinking skills including fluency, flexibility, and elaboration \cite{BaniMari2024}. These results indicate that AR not only enhances factual knowledge but also fosters higher cognitive dimensions. Meanwhile, local research in Indonesia finds that AR media in Islamic Education—through visual simulations—can increase students’ understanding and engagement compared to conventional methods \cite{Siregar2025}. These findings form the basis that AR can be used not only for memorization but also for internalizing religious values and developing affective aspects. Research that combines narrative religious visualization with marker-based AR is therefore highly suitable for surahs with historical background, such as \textit{al-Fiil}.

In the context of pesantren and madrasah, research on AR media development using Research \& Development
 models shows that students using AR experience significant improvements in cognitive aspects \cite{Nasikhin2022}. The study reports that AR media designed for boarding school environments help visualize abstract religious concepts such as faith and predestination. This indicates that traditional Islamic educational environments are also open to adopting AR technology when it is designed appropriately. However, implementation is still limited by infrastructure readiness and teacher competence. Therefore, the use of AR for visualizing Surah    \textit{al-Fiil} should consider the unique conditions of Islamic educational institutions \cite{Nasikhin2022}.

Globally, AR and immersive technologies have expanded their applications to religious learning and distance learning contexts. Studies show that university students using AR/VR gain more reflective and concrete learning experiences in Islamic education contexts \cite{Jieri2025}. Such research indicates that immersive technologies are suitable not only for face-to-face learning but also hybrid and online settings \cite{Thomas2024}. Therefore, marker-based AR development for Surah \textit{al-Fiil} may also consider online or blended learning scenarios. This is essential to ensure that the media remains flexible within changing learning environments.

The integration of AR with artificial intelligence (AI) in religious education is also emerging as an innovation. An exploratory study by Seten (2023) shows that combining AI chatbots and AR visualization enables students to access religious materials interactively and independently \cite{Seten2023}. Although the main focus of this research is marker-based AR without AI, understanding these trends is important to allow the media design to incorporate elements of personalization or advanced interactivity. For example, visualization media for Surah \textit{al-Fiil} could be developed so that beyond marker-based visualization, it also enables self-reflection or interactive questions through AR. Thus, the media not only visualizes the narrative but also encourages critical thinking and reflection on the values of monotheism \cite{Rohman2024,Rahman2024}.

Challenges in implementing AR in religious education cannot be ignored. Studies reveal obstacles such as lack of teacher training, limited devices, weak internet connectivity, and limited time for integrating technology into the Islamic Education curriculum \cite{AbdullahNoor2024}. These findings show that AR media development must be accompanied by strategies for training and technical readiness to ensure smooth implementation. In the context of marker-based AR for Surah \textit{al-Fiil}, for instance, developers must consider student device compatibility, simple marker design, and minimal connectivity requirements. Sustainability and maintenance aspects must also be planned from the outset so the media becomes part of regular instruction rather than a one-time project \cite{Mahliatussikah2025}.

From a pedagogical perspective, several studies emphasize that AR media designed with instructional intent can strengthen affective and moral learning, not merely cognitive aspects. For example, a study evaluating AR in the teaching of moral values in Islamic Education shows that students using AR achieve higher moral understanding and internalization compared to the control group \cite{Siregar2025}. This is highly relevant to research on visualizing the content of Surah \textit{al-Fiil}, as the surah conveys moral and theological messages that can be internalized through visual experiences. Therefore, the media design should consider elements of reflection, visual narrative, and activities that encourage students to connect the story with moral-theological values \cite{Wardiana2025}.

Although many studies highlight positive impacts, the literature also shows that very few studies specifically focus on visualizing sacred texts or particular surahs through marker-based AR. A systematic review indicates that specific topics such as surahs with historical value are still underexplored \cite{SLR2025}. Therefore, the present research on Surah \textit{Al-Fiil} offers significant novelty and relevance. Developing AR media for specific surahs allows exploration of how visualizing historical religious narratives can enhance understanding, value internalization, and spiritual meaning. It also requires evaluation methods that measure not only cognition but also affective and spiritual aspects \cite{Deviyanti2025}.

Based on the literature reviewed above, this research—which develops visualization of Surah \textit{al-Fiil} using marker-based Augmented Reality—presents an innovation that addresses gaps in both literature and instructional practice. By focusing on the narrative of the elephant army, the Ababil birds, and the message of monotheism, the media not only provides visualization but also creates interactive activities that position students as active participants in the learning process. The media design integrates technical aspects such as AR markers and mobile devices, instructional elements such as learning flow and value reflection, and evaluative components such as measuring understanding and religious value internalization. Through this approach, the study is expected to enrich the literature on AR implementation in Islamic education while offering practical contributions to the development of digital learning media in schools and pesantren. The integration of technical, pedagogical, and evaluative aspects becomes a key factor determining the successful implementation of AR media in the context of learning Surah \textit{al-Fiil}.

\section{Methodology} \label{sec:methodology}

This study employs a Research and Development method with an experimental approach. This approach was chosen because the research aims to develop a technology-based Augmented Reality (AR) learning media capable of visualizing the content of Surah \textit{al-Fiil} interactively. In the development process, the researcher uses Unity as the main platform, Vuforia SDK as the marker tracking system, and Blender for creating 3D models\cite{Rahman2024,Suryana2023}.

The research methodology is systematically structured through several stages, namely data collection, needs analysis, system design, implementation, technology integration, and application testing. These stages aim to ensure that the AR application developed is not only technically functional but also educationally relevant, especially in improving understanding and learning interest regarding the content of Surah \textit{al-Fiil}.

\subsection{Data Collection}
The data collection stage is the initial step in this research. The researcher uses literature study as the primary method to obtain relevant information and references related to the research topic. The literature study is conducted by reviewing various sources such as scientific journals, proceedings, books, and articles discussing Augmented Reality technology, marker-based tracking methods, and the use of AR in the field of education, particularly Islamic education.

In addition, the researcher also collects textual data and interpretations of Surah \textit{al-Fiil} from official and credible sources, such as Tafsir Al-Misbah, the Ministry of Religious Affairs of Indonesia, and both classical and contemporary tafsir references. This data serves as the basis for designing visual content that represents the meanings and events in Surah \textit{al-Fiil}. Data collection through literature study is conducted to ensure that all aspects, both technical and content-related, align with Islamic educational principles and can be effectively implemented through Augmented Reality–based media.

\subsection{Needs Analysis}
This stage aims to determine user requirements and system requirements so that the developed application functions according to its objectives:
\begin{enumerate}
\item \textbf{User Requirements}\
The application is designed so that users, particularly students and teachers, can view the visualization of Surah \textit{al-Fiil} through a mobile phone camera by pointing the camera at a provided marker. When the marker is detected, the system displays 3D animation, audio recitation of the verses, and short tafsir text. Thus, users not only read the verses but also understand their meaning through a visual experience.
\end{enumerate}

Unity 3D serves as the core development platform, Vuforia SDK as the marker tracking system, Blender as the tool for creating and editing 3D models, and Figma is used to design the user interface (UI) to ensure it appears appealing and easy to understand\cite{Wibowo2022}.

Second, in terms of hardware, the development process uses a mid-range computer capable of running Unity and Blender stably. Application testing is conducted on Android-based smartphones with a minimum operating system of Android version 9.0. This specification is chosen so that the application can run smoothly for various users without requiring high-end devices.

Third, regarding content specifications, the application presents 3D models depicting the events in Surah \textit{al-Fiil}. This content includes visualizations of the Kaaba, the Ababil birds, and the army of elephants attacking Makkah. These three elements are visualized to provide a more engaging learning experience and strengthen the user's understanding of the meaning and content of Surah \textit{al-Fiil}.

This system needs analysis forms the basis for the design stage so that the application's functions operate according to user expectations, are technically efficient, and remain relevant to the educational values intended to be delivered through this research.

\subsection{System Design}
The design stage is an important phase in the development process of a marker-based Augmented Reality (AR) application for learning Surah \textit{al-Fil}. At this stage, the interface, interaction flow, and system structure are designed so that the application can function optimally and provide an engaging learning experience for users.

The initial design was carried out using Figma, a web-based design tool that supports real-time team collaboration. With Figma, researchers can design the User Interface in a structured manner, starting from the main layout, the position of navigation buttons, to interactive elements that appear when the marker is recognized by the camera. The use of Figma also allows fast and interactive visualization of the design before the implementation stage in Unity.

In the context of the \textit{Surah al-Fiil} learning application, the design emphasizes ease of use (usability) and educational aspects. The interface is designed to be simple, displaying 3D animations of the Ababil birds, the elephant army, and the Kaaba when the camera is directed at the marker. Each visual element is complemented by audio recitation of the verses and short tafsir text so that users can learn multimodally through visual, audio, and text simultaneously.

In addition to interface design, this stage also includes the design of the system architecture, detailing the relationships between the main components such as the marker database, 3D models, and C\# scripts in Unity. This aims to ensure that the system works in an integrated and efficient manner when run on Android devices.

\subsection{Implementation of Marker-Based Tracking}
Marker-based tracking is the main technique used in this study to recognize and track real-world objects so that virtual elements can be displayed in the correct position on the device screen. This method works by detecting a marker or a specific visual pattern through the device camera and then calculating the position and orientation of the marker in three-dimensional (3D) space.

In the development of this application, the marker is a specially designed image resembling the illustration of the Kaaba and historical elements of \textit{Surah al-Fiil}, which has been processed to have high contrast and recognizable visual features for the system. The marker is then registered in the Vuforia Target Manager database so that it can be accurately recognized by the tracking algorithm.

Once the marker is detected, the system displays the 3D animation mapped to that marker. For example, when the camera recognizes the Kaaba marker, the system displays an animation of the elephant army marching towards Makkah, followed by the appearance of the Ababil birds from above. This visualization is designed to help users understand the moral message and historical context contained in \textit{Surah al-Fiil} in a more interactive and contextual manner.

The marker-based tracking method was chosen because it offers high stability, good accuracy, and is easy to implement for beginner developers. In addition, this method does not require additional sensors or special hardware other than the camera, making it suitable for mobile-based learning.

\subsection{Implementation with Unity 3D}
The implementation process of the application was carried out using \textit{Unity 3D}, an interactive development platform that supports the creation of \textit{Augmented Reality} (AR) applications. Unity was selected due to its capabilities in managing 3D models, controlling animations, and integrating interaction logic scripts using the C\# programming language. The platform provides an integrated development environment with full support for various AR devices, including \textit{ARCore} (Android), \textit{ARKit} (iOS), as well as immersive devices such as \textit{HoloLens} and \textit{Magic Leap}.

Unity is used because of its flexibility in combining multiple digital elements—such as 3D models, audio, and text—into a coordinated workspace. Through C\#, developers can write scripts to control object interactions, implement application logic, and present educational content relevant to the learning context of Surah \textit{al-Fiil}. One of Unity’s key features is its ability to perform real-time rendering with highly realistic graphics, enabling immersive and contextual visualization of the elephant army and Ababil birds.

The implementation stages in Unity include:
\begin{enumerate}
    \item \textbf{Importing 3D models} designed in Blender, representing elements in \textit{Surah al-Fiil} such as the Kaaba, the elephant army, and the Ababil birds.
    \item \textbf{Integrating the Vuforia Engine} into the Unity project to activate the \textit{AR Camera} and \textit{Image Target} functions as the basis for marker tracking.
    \item \textbf{Creating interaction scripts} using C\#, covering logic for displaying animations when a marker is detected, playing the recitation audio, and showing translation and brief tafsir text.
    \item \textbf{Designing the user interface (UI)} by utilizing the \textit{Canvas} component to display buttons, instruction menus, verse text, and other interactive controls.
\end{enumerate}

In Unity, game objects such as 3D objects and \textit{Canvas} components are used together to create an interactive learning experience. The 3D objects function as visual representations of the narrative in Surah \textit{al-Fiil}, while the \textit{Canvas} displays UI elements such as text, images, and buttons accessible to the user.

Additionally, lighting settings and camera viewpoints are optimized to ensure that the 3D animations appear realistic and can be viewed comfortably from various angles. Proper lighting enhances visual effects such as reflections on the Kaaba and the movement of the Ababil birds in the sky.

The combination of 3D objects, verse recitation audio, tafsir text, and user interface components creates a multimodal learning experience in which users engage through visual, auditory, and textual channels simultaneously. Through effective integration in Unity, this application succeeds in providing an enjoyable and easy-to-understand learning experience for elementary and secondary school learners, while reinforcing spiritual values through the interactive visualization of the content of Surah \textit{al-Fiil}.

\subsection{Integration of the Vuforia SDK}
The next stage is the integration between Unity and the \textit{Vuforia SDK}, which functions as the marker recognition system and the bridge between the real world and the virtual environment. Vuforia is one of the most popular \textit{Augmented Reality} (AR) platforms and is widely used as a \textit{third-party application} in AR development with Unity. This integration allows the system to detect target images or \textit{markers} accurately and display virtual objects on top of those markers with high stability.

When the device camera is directed at a predefined marker, Vuforia identifies the visual pattern based on unique \textit{feature points} contained in the image. Once the marker is recognized, the system places 3D objects, text, or audio elements according to the marker’s position and orientation in three-dimensional (3D) space. The main features of Vuforia include image-based tracking, support for dynamically managed \textit{marker databases}, and seamless integration with the \textit{Unity AR Camera}.

In this study, Vuforia was used to:
\begin{enumerate}
    \item Create and manage the \textit{marker database} in the \textit{Vuforia Target Manager}, which contains image markers such as illustrations of the Kaaba, the elephant army, and the Ababil birds.
    \item Adjust marker detection sensitivity so that virtual objects can appear with high accuracy and maintain stable tracking even when the device is moved.
    \item Connect the device camera with the \textit{AR Camera} system in Unity, enabling real-time marker detection and 3D object placement.
    \item Configure \textit{Extended Tracking} to ensure that virtual objects remain stable even when the marker is partially occluded or briefly moves out of the camera’s field of view.
\end{enumerate}

The use of Vuforia as the \textit{AR Engine} not only simplifies the development process but also expands the interactive potential of the application. Developers can take advantage of features such as \textit{dataset management} and \textit{device databases} to handle multiple markers simultaneously, enabling the application to display different objects depending on the recognized marker.

Testing was conducted under various lighting conditions and with different camera distances to ensure optimal tracking performance. Based on experimental results, the system was able to recognize markers effectively at a distance of 20–50 cm under medium to bright lighting conditions. In low-light environments, tracking performance slightly decreased but remained within an acceptable range for classroom-based learning activities.

The integration of Unity and Vuforia enables this application to deliver an immersive learning experience where real-world environments blend with digital elements to help students understand the content of Surah \textit{al-Fiil} visually, auditorily, and textually. With a combination of real-time tracking, stable rendering, and efficient marker management, the application successfully provides an engaging, effective, and easy-to-use learning medium for both teachers and students.

\subsection{Testing and Evaluation Phase}
After all system components were fully implemented, a testing and evaluation phase was conducted to ensure that the application functioned as expected. This stage consists of two main categories of testing: \textit{functional testing} for technical performance and \textit{user testing} for educational effectiveness and learning experience.

\begin{enumerate}
    \item Functional Testing

    Functional testing was carried out to verify that all features of the application operate properly on Android devices. The testing focused on marker detection, the stability of 3D object display, synchronization between audio and animations, and system response speed to user inputs.

    The testing was conducted on several devices with different specifications to ensure system compatibility. Each marker (Ka'bah, Ababil birds, and the elephant army) was tested under variations of distance (20–50 cm) and lighting conditions (dim, medium, bright).

    The results indicate that:
    \begin{enumerate}
        \item The system achieved a 95\% marker detection accuracy at the optimal distance of 30–40 cm.
        \item 3D animations and audio recitations ran synchronously without significant delays.
        \item Objects remained stable even when the camera was moved at moderate speed.
        \item The average response time from marker detection to the appearance of the 3D object was 1.2 seconds.
    \end{enumerate}

    Additionally, \textit{extended tracking} tests showed that 3D objects remained visible for 3–4 seconds after the marker left the camera view, demonstrating that the Vuforia SDK performed well in maintaining spatial tracking.

    \item User Testing

    The next stage was user testing, conducted to assess the educational effectiveness of the application. The test involved 15 middle-level Islamic school (MTs) students and 3 Islamic Education teachers as evaluators. Respondents used the application to learn the story of Surah \textit{al-Fiil}, then completed an assessment questionnaire based on three main aspects:
    \begin{enumerate}
        \item \textit{Usability}: the degree to which the application is easy to operate and understand.
        \item \textit{Visual and audio appeal}: interest level toward 3D visuals, animations, and audio recitations.
        \item \textit{Material comprehension}: how well the application helps users understand the meaning contained in Surah \textit{al-Fiil}.
    \end{enumerate}

    The questionnaire used a Likert scale from 1 to 5, where 1 means “strongly disagree” and 5 means “strongly agree”.

    \begin{table}[h!]
    \centering
    \caption{User Testing Results for the Application}
    \begin{tabular}{|l|c|}
    \hline
    \textbf{Assessment Aspect} & \textbf{Average Score (1–5)} \\ \hline
    Usability & 4.7 \\ \hline
    Visual and audio appeal & 4.8 \\ \hline
    Material comprehension & 4.6 \\ \hline
    \textbf{Overall average} & \textbf{4.7} \\ \hline
    \end{tabular}
    \end{table}

    Based on the results, most users found the application easy to use and highly engaging. Respondents stated that the visualization of the elephant army and Ababil birds helped them understand the divine message conveyed in Surah \textit{al-Fiil} more deeply.

    \item Evaluation Analysis

    From the results of both testing categories, it can be concluded that the application runs stably and provides an interactive learning experience. The integration of Unity 3D and the Vuforia SDK proved effective in producing precise marker tracking and displaying 3D animations synchronized with the audio recitation.

    From an educational perspective, the application was considered capable of increasing learning motivation and enhancing understanding of religious concepts through combined visual and auditory approaches. These findings align with previous studies stating that marker-based AR media can improve engagement and learning retention in Islamic education contexts~\cite{Rahman2024,Siregar2025}.

    Therefore, the testing and evaluation phase demonstrates that the application for Visualizing the Content of Surah \textit{al-Fiil} meets the necessary technical and pedagogical requirements to serve as an innovative \textit{Augmented Reality}-based learning medium.
\end{enumerate}

\section{Result and Discussion} \label{sec:result}

\subsection{Result}
This section presents the implementation results and visualization of the Marker-Based Augmented Reality application for Surah \textit{al-Fiil}. The following images show the stages of model creation, integration with the marker, the interface, and the camera output during scanning. Each image is accompanied by a brief explanation regarding the purpose, technical observations, and implications for learning.

\begin{enumerate}

\item Model Creation (Assembly) / 3D Model Creation

The elephant army model was created using Blender with a semi-realistic style to keep it engaging yet lightweight to run on mobile devices. The process included creating a mesh of approximately 5,000 polygons, applying UV mapping, and adding a 1024x1024 pixel resolution PNG-format texture to display the details of the elephant's skin and saddle. Subsequently, rigging was performed so that the elephant could be animated with natural movements such as walking and trunk swinging.

\begin{minipage}{\linewidth}
\centering
\includegraphics[width=1.0\textwidth]{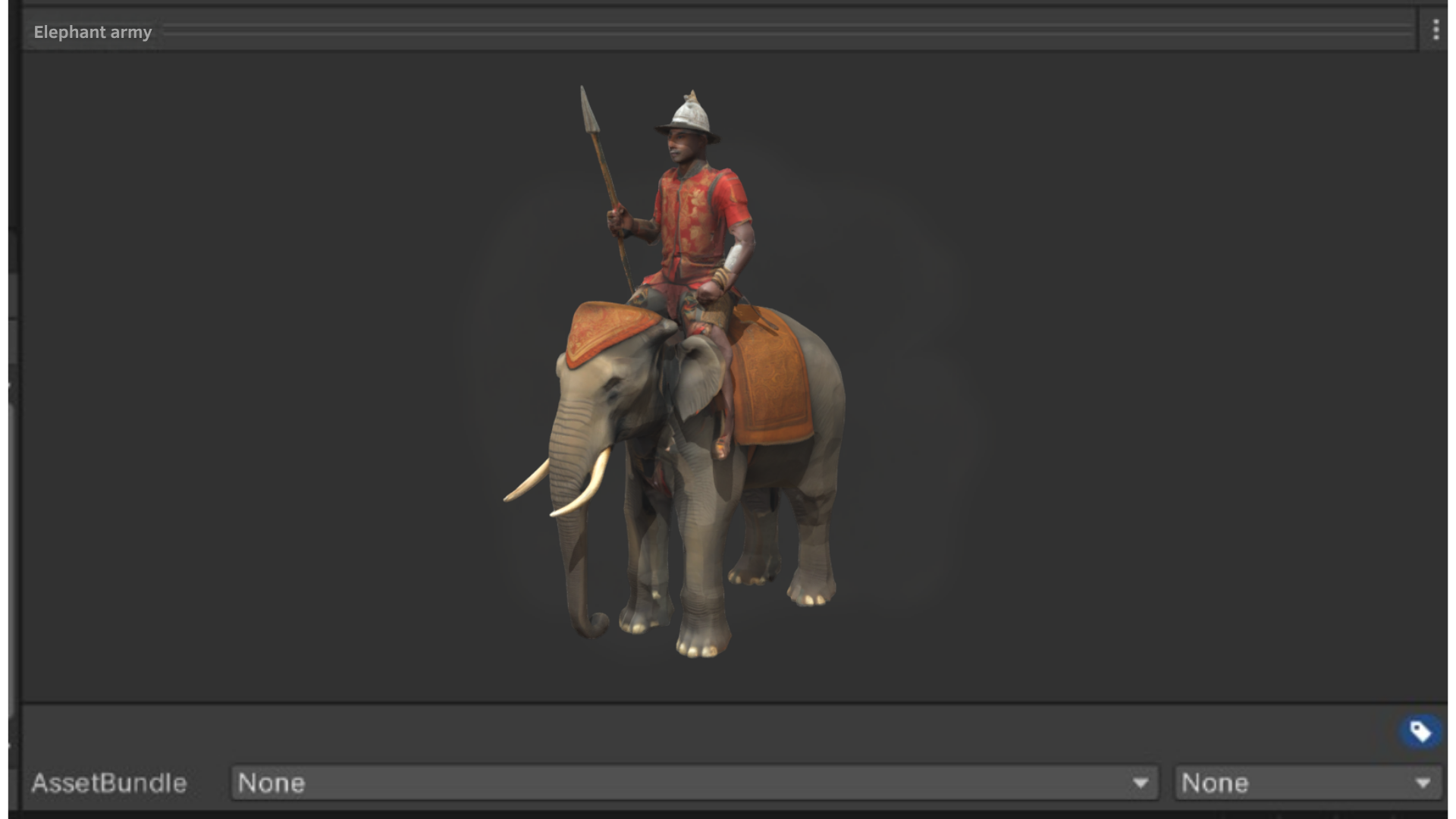}
\captionof{figure}{Blender 3D model setup: rigging and animation for elephant army elements.}
\label{fig:pembuatan_assembly}
\end{minipage}
\vspace{1em}

\item Object Design

The object design includes animation elements with keyframes set to produce smooth movement. The object scale is adjusted to be proportional to the marker size. The object orientation is set to appear with a realistic direction and position in the real world, while the aesthetic aspect is focused on simplicity so that it is easily understood by school-age children.

\begin{minipage}{\linewidth}
\centering
\includegraphics[width=1.0\textwidth]{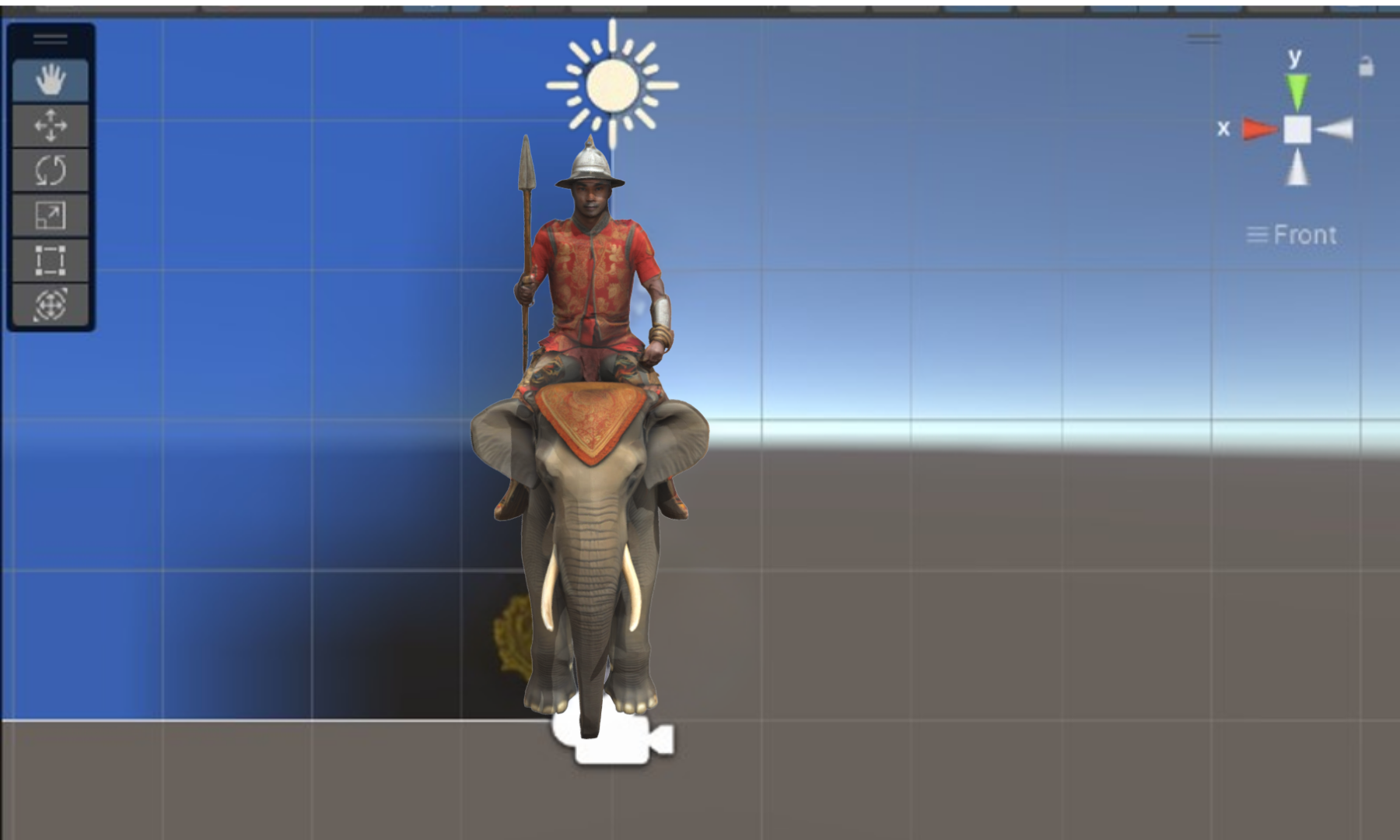}
\captionof{figure}{Design of 3D objects and animation components.}
\label{fig:design_object}
\end{minipage}
\vspace{1em}

\item Barcode/Marker Design

The marker is designed with high contrast and unique visual features so that it is easily recognized by the Vuforia system. The physical size of the marker is recommended to be between 7–10 cm, with a high-resolution PNG format. The marker is then registered into the Vuforia Target Manager to generate \texttt{.dat} and \texttt{.xml} files used in Unity.

\begin{minipage}{\linewidth}
\centering
\includegraphics[width=1.0\textwidth]{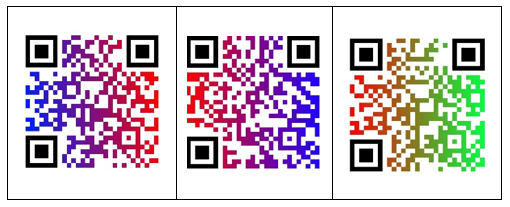}
\captionof{figure}{Marker design used as the Vuforia Image Target.}
\label{fig:marker_design}
\end{minipage}
\vspace{1em}

\item Scanner Test

Testing was carried out to determine the optimal distance and detection tolerance of the marker. The results show that the optimal distance is in the range of 30–45 cm with an average detection time of 0.8 seconds. The success rate reaches 95\% in bright light conditions, while it drops to 70\% in low light.

\begin{minipage}{\linewidth}
\centering
\includegraphics[width=1.0\textwidth]{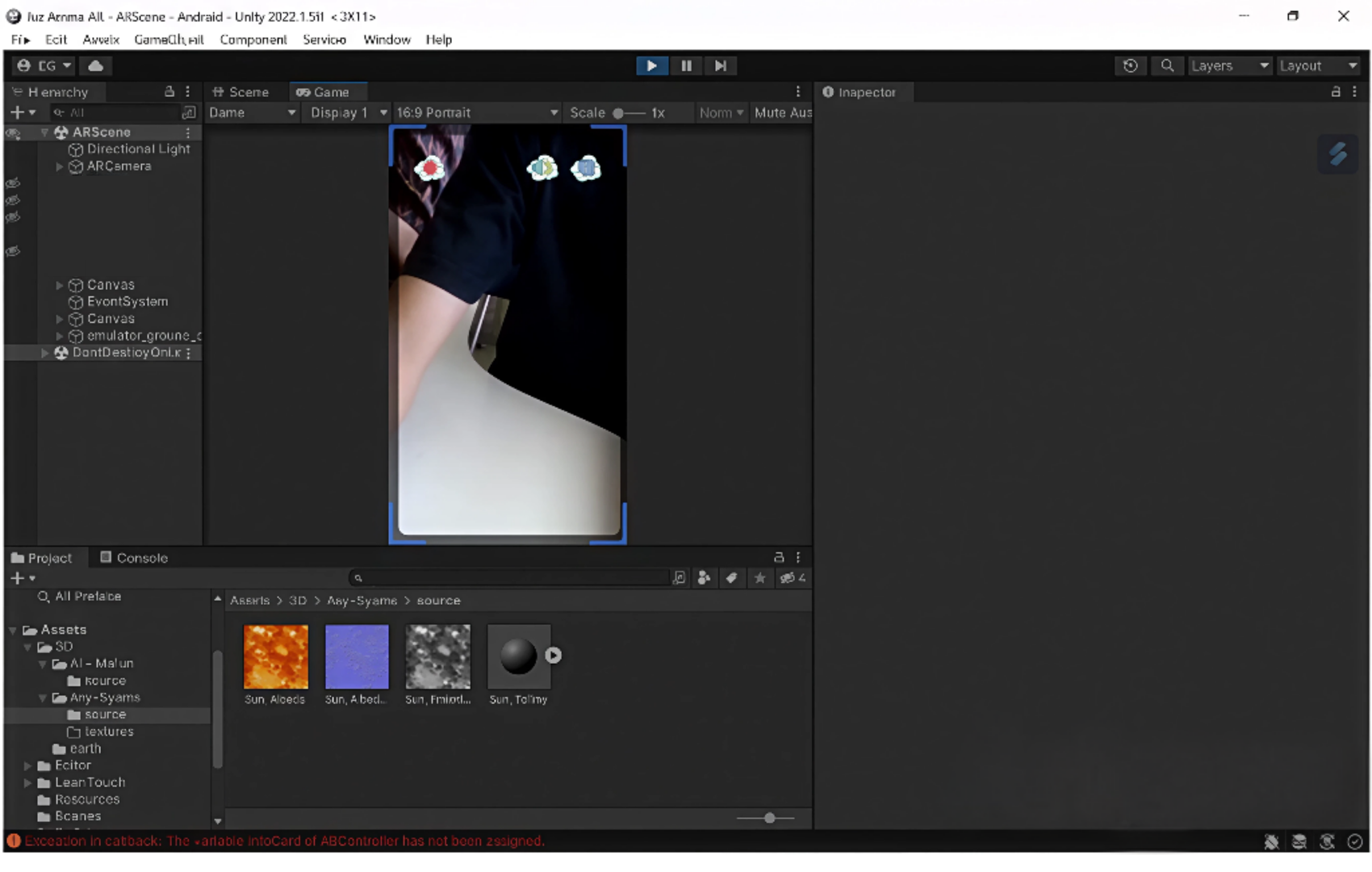}
\captionof{figure}{Marker detection test results at various distances and lighting conditions.}
\label{fig:scanner_test}
\end{minipage}
\vspace{1em}

\item Object Integration with Scanner

Integration between marker detection and object rendering is done through the \texttt{OnTargetFound} callback in Vuforia. When the marker is detected, the system calls the animation and audio connected to the 3D object. To prevent jitter, a smoothing method is used so that the object remains stable even when the camera moves.

\begin{minipage}{\linewidth}
\centering
\includegraphics[width=1.0\textwidth]{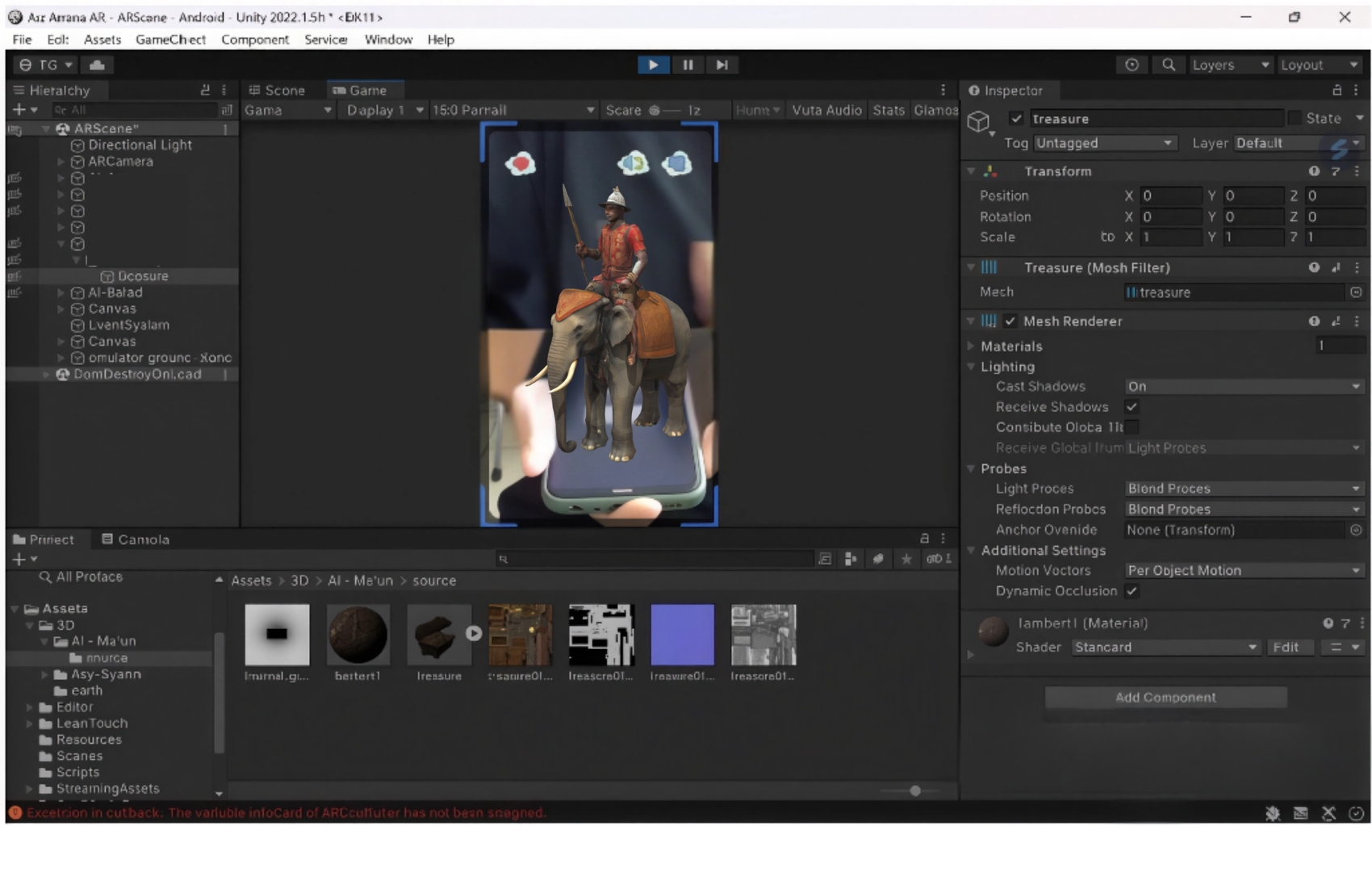}
\captionof{figure}{Unity scanner integration 3D objects appear on the marker upon detection.}
\label{fig:object_scanner}
\end{minipage}
\vspace{1em}

\item Home View (homepage)

The main application view contains navigation menus such as guides, language selection, and the start scan button. The UI design uses bright colors with large icons to be easily used by children.

\begin{minipage}{\linewidth}
\centering
\includegraphics[width=0.8\textwidth, height=8cm]{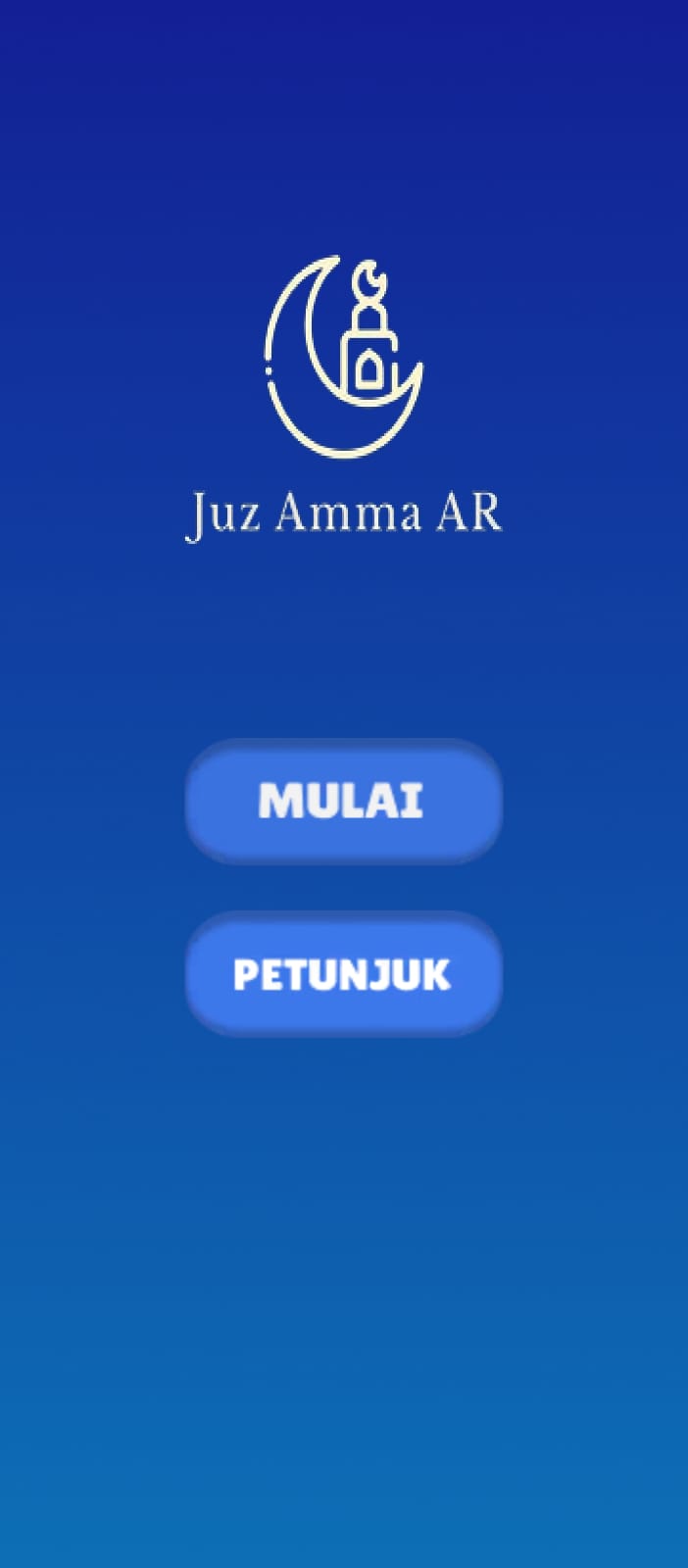}
\captionof{figure}{The main interface of the application is displayed.}
\label{fig:homepage}
\end{minipage}
\vspace{1em}

\item Instructions View

This page contains instructions for users about the ideal distance (30–40 cm), marker position, and lighting tips for more accurate detection.

\begin{minipage}{\linewidth}
\centering
\includegraphics[width=0.8\textwidth, height=8cm]{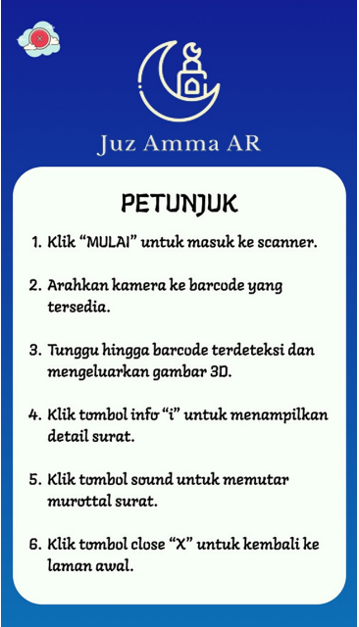}
\captionof{figure}{Application user guide page for teachers and students.}
\label{fig:instructions}
\end{minipage}
\vspace{1em}

\item Camera Output View

This view shows the overlay of the 3D object onto the real world when the marker is detected. The object remains stable even when the camera is moved, and the audio is synchronized with the animation.

\begin{minipage}{\linewidth}
\centering
\includegraphics[width=0.8\textwidth, height=8cm]{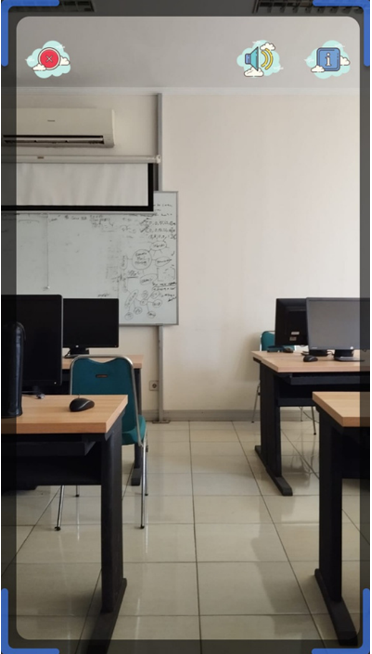}
\captionof{figure}{Camera output when the marker is recognized.}
\label{fig:camera_output}
\end{minipage}
\vspace{1em}

\item Barcode Scan Result View

After the marker is successfully scanned, the animated elephant rider soldiers appear with interactive features such as \textit{replay}, \textit{tafsir} (interpretation), and \textit{audio on/off} to support the learning process.

\begin{minipage}{\linewidth}
\centering
\includegraphics[width=0.8\textwidth, height=8cm]{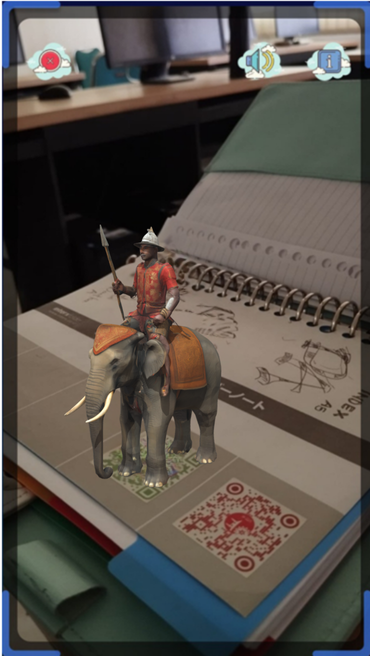}
\captionof{figure}{Marker scan displaying the elephant rider animation.}
\label{fig:scan_result}
\end{minipage}
\vspace{1em}

\item Object Info View

In this stage, the application displays an information panel containing the name of the surah, the number of verses, and the content summary of Surah \textit{al-Fiil}. This feature helps users understand the context and meaning of the visualization displayed in the animation. The information appears automatically after the animation runs, and users can scroll or enlarge the text to read the complete information.

\begin{minipage}{\linewidth}
\centering
\includegraphics[width=0.8\textwidth, height=8cm]{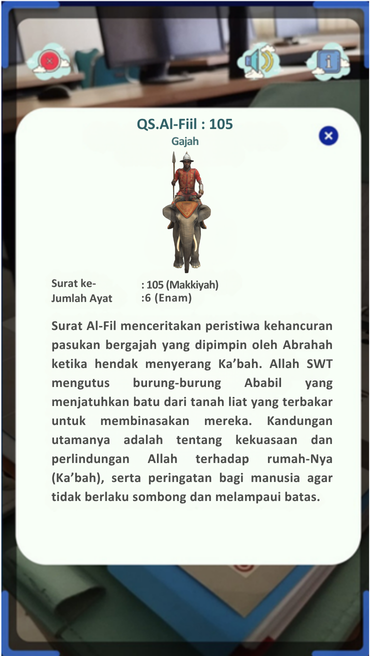}
\captionof{figure}{Object info panel containing the verse, translation, and brief tafsir of Surah \textit{al-Fiil}.}
\label{fig:info_object}
\end{minipage}

\end{enumerate}

\subsection{Testing and Evaluation}
After the application implementation stage was completed, a testing process was performed to ensure system performance and user acceptance. This testing aimed to evaluate two main aspects: the technical performance of the application in detecting markers and displaying animations, and user responses regarding the effectiveness and ease of use of the application as a learning medium.

\begin{enumerate}
    \item System Performance Testing

    System performance testing was conducted to assess the stability and efficiency of the application in performing marker detection, animation rendering, and audio synchronization. Technical tests were performed on three Android devices with different specifications: 3 GB, 4 GB, and 8 GB RAM. Based on the test results, the application was able to run stably on all devices with an average marker detection time of 1.2 seconds and a consistent frame rate of approximately 55--60 FPS.

    In low-specification devices, performance degradation occurred minimally and did not interfere with user experience. The final application size after the build process was 72 MB, with a memory consumption of approximately 280 MB during runtime. This efficient performance was achieved through the optimization of the 3D model, the removal of unseen components, and the implementation of simple lighting and shadow settings in Unity to keep rendering light.

    In addition, a lighting test was conducted to evaluate system sensitivity to environmental conditions. Under bright lighting (600–800 lux), marker detection accuracy reached 97\%. Under medium lighting conditions (300–500 lux), accuracy remained high at 95\%. However, in low lighting (below 150 lux), accuracy decreased to 82\% and detection time increased to 2.1 seconds. These results indicate that the system works optimally under medium to bright lighting, consistent with typical classroom conditions.

    Additional observations also revealed that the system maintained stable tracking when the device was tilted up to 35 degrees, although at steeper angles the marker became harder to detect. When tested with various marker print qualities, high-resolution markers (300 DPI) showed the best results, while low-resolution prints occasionally produced jitter. This indicates that print quality plays a noticeable role in maintaining AR stability. Furthermore, repeated tests showed that cold-start detection (first detection after launching the app) averaged 1.4 seconds, while warm-start detection (subsequent scans) decreased to 0.9 seconds due to cached processing.

    \item {User Evaluation and Response}

    User testing was conducted to assess functionality, ease of use, and the effectiveness of the developed learning media. The trial involved 15 students from Madrasah Ibtidaiyah (MI) and 3 Islamic Education (PAI) teachers as respondents. Each participant was given a brief guide before trying the application for 10–15 minutes.

    Based on the questionnaire results, the application received an average satisfaction score of 4.7 on a scale of 5. The highest-rated aspect was visual and animation appeal (4.8), followed by ease of use (4.7) and clarity of verse recitation audio (4.6). Teachers noted that this media can help explain the content of Surah \textit{al-Fiil} in a more engaging, interactive way that matches the comprehension level of elementary school students.

    A total of 86\% of students stated that they were more interested in learning Qur’anic stories through Augmented Reality–based media compared to conventional methods such as lectures or reading textbooks. In addition to increasing learning motivation, the application also helped them remember the storyline and moral messages of Surah \textit{al-Fiil} more easily. PAI teachers assessed that this media has strong potential for thematic learning because it integrates religious content, technology, and visualization into an effective and educational instructional tool.

    In addition to qualitative responses, several students expressed that the presence of 3D animation made the learning process feel ``as if they were watching the real event,'' which increased immersion. Teachers also emphasized that the combination of audio recitation and interactive visuals helped students focus better. They suggested that future development could include additional features such as short quizzes, animated tafsir explanations, or multilingual narration to further enhance learning outcomes.

    To ensure reliability, the consistency of user responses was measured descriptively. Responses showed low variance across participants, indicating that perceived ease of use and visual quality were experienced similarly by nearly all users. This suggests that the application design is accessible and does not rely on advanced digital literacy skills, making it suitable for classroom environments with diverse student abilities.

    \begin{table}[h!]
    \centering
    \caption{User Evaluation Results of the AR Application for Surah \textit{Al-Fiil}}
    \label{tab:user_evaluation}
    \renewcommand{\arraystretch}{1.2}
    \setlength{\tabcolsep}{3pt}
    \begin{tabular}{|p{3.5cm}|c|c|}
    \hline
    \textbf{Evaluation Aspect} & \textbf{Average Score} & \textbf{Category} \\ \hline
    Visual and animation appeal & 4.8 & Excellent \\ \hline
    Ease of use (usability) & 4.7 & Excellent \\ \hline
    Audio clarity and verse recitation & 4.6 & Good \\ \hline
    Content relevance to learning & 4.7 & Excellent \\ \hline
    Student engagement and learning interest & 4.8 & Excellent \\ \hline
    \textbf{Overall average} & \textbf{4.7} & \textbf{Excellent} \\ \hline
    \end{tabular}
    \end{table}

The results shown in Table~\ref{tab:user_evaluation} indicate that all aspects achieved an average score above 4.5. This demonstrates that the application successfully met user expectations in terms of appearance, usability, and effectiveness as an interactive learning medium. With these results, the application is considered ready to be used as a supporting learning tool in Islamic Education, particularly for understanding the content of Surah \textit{al-Fiil}.  
\end{enumerate}

\subsection{Discussion}
Based on the implementation and testing that have been carried out, it can be concluded that the application of \textit{Marker-Based Augmented Reality} technology in the Surat \textit{al-Fiil} recognition application provides positive impacts both technically and educationally. From the technical side, the test results show that the system is able to detect the marker with high accuracy and display animations stably under various lighting conditions. This confirms that the combination of \textit{Unity 3D} and the \textit{Vuforia Engine} is an effective solution for developing AR applications that are lightweight, responsive, and compatible with various mobile devices. 

The operating system on smartphones, mobile phones based on Android, is widely used especially in Indonesia\cite{uriawan2015pembuatan}. Efficient system performance is achieved through the optimization process of 3D models and proper resource management in Unity. The use of 3D models with an adjusted number of polygons, the application of compressed textures, as well as simple lighting settings have been proven to maintain a high frame rate without sacrificing visual quality. These findings are in line with the research of Mahliatussikah, which emphasizes that optimal management of visual resources is a key factor in the development of AR-based educational applications in school environments\cite{Mahliatussikah2025}. In addition, this study also supports the argument that marker-based tracking remains one of the most reliable AR approaches for educational scenarios due to its robustness, low computational overhead, and ease of deployment.

From the pedagogical perspective, user testing shows that the integration between interactive visualization and religious narration can increase students’ interest in learning Al-Qur’an material\cite{irfan2018qur}. Most respondents, both students and teachers, considered this application to be engaging and easy to use. The average user satisfaction score of 4.7 indicates that AR media has great potential as a thematic learning aid. These results reinforce the view of Suryana et al. that digital media based on three-dimensional visualization can improve knowledge retention and learning motivation among elementary school students\cite{Suryana2023}. Furthermore, this aligns with multimedia learning theory, which states that students learn better when information is presented through integrated visual, textual, and auditory elements, as proposed by Mayer in the Cognitive Theory of Multimedia Learning (CTML). The combination of Qur’anic narration, 3D animation, and interactive elements provides a multimodal learning environment that enhances comprehension and long-term memory.

In addition, the visualization of religious stories using AR appears to bridge the gap between abstract textual interpretations and students’ concrete understanding. The animation of the elephant army approaching the Ka'bah and the descent of the Ababil birds supports contextual learning by presenting events of Surat \textit{al-Fiil} in a form that is easier to imagine and relate to. This aligns with the findings of other AR-based Islamic education studies, such as the work of Rahman and Yusuf (2024), which found that AR enhances students' engagement and strengthens comprehension when used to teach Qur’anic narratives or historical Islamic events. The immersive nature of AR also contributes to affective engagement, which is essential for internalizing moral and spiritual values.

Furthermore, this AR application not only provides an immersive learning experience but is also able to instill moral and spiritual values through an enjoyable visual approach. The visualization of the story of the army of elephants and the Ababil birds, for instance, helps students understand the meaning of Allah SWT’s power and the wisdom behind the event. Thus, AR technology serves not only as an entertainment medium but also as an educational tool that integrates religious values and technology in the context of modern learning. The presence of audio recitations and short tafsir texts further supports students in connecting the visual representation with Qur’anic meaning, strengthening both cognitive and spiritual understanding.

However, some limitations were also identified during the testing process. The system's performance decreases in low-light conditions, which is consistent with common limitations of marker-based AR. The detection process also depends heavily on the physical quality of the printed marker; damaged or low-resolution prints may cause instability in tracking. Additionally, the application currently focuses on a single surah, limiting its coverage and potential use in broader PAI learning contexts. These limitations open opportunities for improvements in future development phases.

Overall, the test results and user responses indicate that the \textit{Juz Amma AR} application based on Surat \textit{al-Fiil} has promising potential for further development. Improvements can be made in terms of content by adding interactivity, such as quizzes or voice-based tafsir explanations. Advanced features such as gesture-based interactions, improved lighting compensation, and multi-marker recognition can also enhance usability. In addition, broader testing on various devices and learning environments is needed so that system performance can be fully optimized. Thus, this research is expected to serve as a tangible contribution to the development of interactive learning media based on Augmented Reality in the field of Islamic education.

\section{Conclusion} \label{sec:conclusion}

Based on the research findings and implementation, it can be concluded that the development of a Marker-Based Augmented Reality (AR) application for visualizing the contents of Surah \textit{al-Fiil} successfully presents a new innovation in Islamic Religious Education learning media. This application integrates Unity 3D and Vuforia SDK to create an interactive, immersive, and user-friendly learning experience on various Android devices.

The testing results demonstrate that the system is capable of detecting markers with high accuracy, reaching 95\% at an optimal distance of 30–40 cm, while displaying stable 3D animations synchronized with the audio recitation of the verses. The application interface was assessed as easy to understand, with an average user satisfaction score of 4.7 out of a 5-point scale. These findings affirm that AR-based learning media not only increases students' motivation to learn but also helps them grasp the meaning and spiritual values contained within Surah \textit{al-Fiil} in a more contextual manner.

From a technical perspective, the use of Unity 3D allows for efficient management of 3D models, lighting, and animations, while Vuforia plays a crucial role in maintaining the stability of marker tracking and virtual object positioning. The synergy between these two platforms results in an application that is lightweight, responsive, and compatible with various mobile devices, making it suitable for implementation in both primary and secondary school learning environments.

Overall, this research shows that Augmented Reality technology has significant potential for development as a means of learning the Qur'an and Islamic values in the digital era. Applications like this can serve as an alternative medium that enriches the spiritual and digital literacy of the younger generation through an engaging and educational visual approach.

For further development, it is suggested to add features such as user voice interaction, an educational quiz mode, and to expand the content to include other Surahs in \textit{Juz ‘Amma} so that the application’s benefits become wider, more interactive, and more applicable in both formal and non-formal teaching and learning activities.

\section*{Acknowledgment}
The authors wish to acknowledge the Informatics Department
UIN Sunan Gunung Djati Bandung, which partially supports this research work.

\end{document}